\input harvmac
\input epsf

\overfullrule=0pt
\abovedisplayskip=12pt plus 3pt minus 3pt
\belowdisplayskip=12pt plus 3pt minus 3pt
\sequentialequations

\def\tphi{{\tilde\phi}}

\lref\qt{Q. Tarantino, {\it ``Pulp Fiction''}, Miramax
Pictures (1994).}
\lref\orien{A. Sagnotti, {\it ``Open strings and their symmetry 
groups''}, in: Non-perturbative Quantum Field Theory, Cargese 1987,
eds. G.Mack et. al. (Pergamon Press 1988)\semi 
P. Horava, {\it ``Background duality of open string models''}, 
Phys. Lett. {\bf B231} (1989), 251.}
\lref\dlp{J. Dai, R.G. Leigh and J. Polchinski, {\it ``New connections
between string theories''}, Mod. Phys. Lett., {\bf A4} (1989), 2073.}
\lref\dufflu{M.J. Duff and J.X. Lu, {\it ``Stringy solitons''},
hep-th/9412184, Phys. Rep. {\bf 259} (1995), 213.}
\lref\stromopen{A. Strominger, {\it ``Open p-branes''}, hep-th/9512059,
Phys. Lett. {\bf B383} (1996), 44.}
\lref\townopen{P.K. Townsend, {\it ``D-branes from M-branes''},
hep-th/9512062, Phys. Lett. {\bf B373} (1996), 68.}
\lref\schwarzpq{J.H. Schwarz, {\it ``An SL(2,Z) multiplet of type IIB
superstrings''}, hep-th/9508143, Phys. Lett. {\bf B360} (1995), 13.}
\lref\romans{L.J. Romans, {\it ``Massive N=2A supergravity in
ten-dimensions''}, Phys. Lett. 169B (1986), 374.}
\lref\pold{J. Polchinski, {\it ``D-branes and Ramond-Ramond charges''},
hep-th/9510017, Phys. Rev. Lett. {\bf 75} (1995), 47.}
\lref\polwit{J. Polchinski and E. Witten, {\it ``Evidence for 
heterotic-type I string duality''}, hep-th/9510169,
Nucl. Phys. {\bf B460} (1996), 525.}
\lref\morsei{D.R. Morrison and N. Seiberg, {\it ``Extremal transitions
and five-dimensional supersymmetric field theories''},
hep-th/9609070, Nucl. Phys. {\bf B483} (1997), 229.}
\lref\senf{A. Sen, {\it ``F-theory and orientifolds''}, 
hep-th/9605150, Nucl.  Phys. {\bf B475} (1996), 562.}
\lref\vafaf{C. Vafa, {\it ``Evidence for F-theory''}, hep-th/9602022,
Nucl. Phys. {\bf B469} (1996), 403.}
\lref\dmf{K. Dasgupta and S. Mukhi, {\it ``F-theory at constant 
coupling''}, hep-th/9606044, Phys. Lett. {\bf B385} (1996), 125.}
\lref\johan{A. Johansen, {\it ``A comment on BPS states in F-theory in
8 dimensions''}, hep-th/9608186, Phys. Lett. {\bf B395} (1997), 36.}
\lref\gabz{M. Gaberdiel and B. Zwiebach, {\it ``Exceptional
groups from open strings''}, hep-th/9709013.}
\lref\dmm{K. Dasgupta and S. Mukhi, {\it ``Orbifolds of M-theory''},
hep-th/9512196, Nucl. Phys. {\bf B465} (1996), 399.}
\lref\witfive{ E. Witten, {\it ``Five branes and M-theory on an 
orbifold''}, hep-th/9512219, Nucl. Phys. {\bf B463} (1996), 383.}
\lref\seibthree{N. Seiberg, {\it ``IR dynamics on branes and 
space-time geometry''}, hep-th/9606017, Phys.Lett. {\bf B384} (1996), 81.}
\lref\seiwitthree{N. Seiberg and E. Witten, {\it ``Gauge dynamics and
compactification to three dimensions''}, hep-th/9607163.}
\lref\bsv{M. Bershadsky, V. Sadov and C. Vafa, {\it ``D-branes and 
topological field theories''}, hep-th/9611222, Nucl. Phys. {\bf B463}
(1996), 420.}
\lref\witflux{E. Witten, {\it ``Flux quantization in M-theory and the 
effective action''}, hep-th/9609122.}
\lref\djm{K. Dasgupta, D. Jatkar and S. Mukhi, {\it ``Gravitational
couplings and $Z_2$ orientifolds''}, hep-th/9707224.}
\lref\svw{S. Sethi, C. Vafa and E. Witten, {\it ``Constraints 
on low-dimensional string compactifications''}, hep-th/9606122,
Nucl. Phys. {\bf B480} (1996), 213.}
\lref\dmn{K. Dasgupta and S. Mukhi, {\it ``A note on low-dimensional 
string compactifications''}, hep-th/9612188, Phys. Lett. {\bf B398}
(1997), 285.}
\lref\ghm{M.B. Green, J.A. Harvey and G. Moore, {\it ``I-brane inflow and 
anomalous couplings on D-branes''}, hep-th/9605033,
Class. Quant. Grav. {\bf 14}(1997) 47.}
\lref\witvar{E. Witten, {\it ``String theory dynamics in various
dimensions''}, hep-th/9503124, Nucl. Phys. {\bf B443} (1995), 85.}
\lref\townm{P.K. Townsend, {\it ``The eleven-dimensional supermembrane 
revisited''}, hep-th/9501068, Phys. Lett. {\bf B350} (1995), 184.}
\lref\horwit{P. Horava and E. Witten, {\it ``Heterotic and type I 
string dynamics from eleven-dimensions''}, hep-th/9510209,
Nucl. Phys. {\bf B460} (1996), 506.} 
\lref\salamsez{Abdus Salam and E. Sezgin (eds), {\it ``Supergravities in
diverse dimensions''}, World Scientific, Singapore (1989).}
\lref\sensix{A. Sen, {\it ``Strong coupling dynamics of branes from
M-theory''}, hep-th/9708002.}
\lref\bersix{M. Berkooz, R.G. Leigh, J. Polchinski, J.H. Schwarz,
N. Seiberg, E. Witten {\it ``Anomalies, dualities and topology of
D=6, N=1 superstring vacua''},  hep-th/9605184, Nucl. Phys. {\bf B475}
(1996), 115.}

{\nopagenumbers
\Title{\vtop{\hbox{hep-th/9710004}
\hbox{TIFR/TH/97-54}}}
{\vtop{\centerline{Orientifolds:}
\medskip
\centerline{The Unique Personality Of Each Spacetime Dimension}}}%
\footnote{}{Based on talks delivered at the Workshop on Frontiers 
in Field Theory, Quantum Gravity and String Theory, Puri (December
1996) and at the joint Paris-Rome-Utrecht-Heraklion-Copenhagen
meeting, Ecole Normale Superieure, Paris (August 1997).}

\centerline{Sunil Mukhi\foot{E-mail: mukhi@theory.tifr.res.in}}
\vskip 4pt
\centerline{\it Tata Institute of Fundamental Research,}
\centerline{\it Homi Bhabha Rd, Bombay 400 005, 
India}
\ \medskip
\centerline{ABSTRACT}

I review some recent developments in our understanding of the dynamics
of $Z_2$ orientifolds of type IIA and IIB strings and
M-theory. Interesting physical phenomena that occur in each spacetime
dimension from 10 to 1 are summarized, along with some relationships
to nonperturbative effects and to M- and F-theory. The conceptual
aspects that are highlighted are: (i) orientifolds of M-theory, (ii)
exceptional gauge symmetry from open strings, (iii) certain
similarities between branes and orientifold planes. Some comments are
also made on disconnected components of orientifold moduli space.
\ \vfill 
\leftline{September 1997}
\eject}
\ftno=0

\centerline{\it ``Personality goes a long way''\refs{\qt}.}

\newsec{Introduction: Orientifolds}
An {\it orientifold} of an oriented, closed-string theory is obtained
by taking the quotient of the theory by the action of a discrete
symmetry group $\Gamma$ that includes orientation reversal of the
string\refs{\orien,\dlp}. After taking the quotient, one has a theory
of unoriented closed and open strings. In this article I will restrict
my attention to cases where $\Gamma=Z_2$, so the single nontrivial
element is $\Omega \cdot I$ where $I$ is some geometric symmetry that
squares to 1.

The unoriented closed strings are simply what is left of the original
oriented strings after their two possible orientations have been
identified. Open strings arise in a way analogous to the appearance
of ``twisted sectors'' in string orbifolds (though this analogy should
not be carried too far). The presence of open strings is required to
cancel anomalies and tadpoles, and is therefore crucial to the
consistency of the theory, much as orbifold twisted sectors are
necessary to maintain modular invariance.

The elements of the discrete group $\Gamma$ typically have fixed
points. More precisely, the elements of $\Gamma$ usually act on some
compactified factor of the full spacetime, and have fixed points
there. These points are then fixed hyperplanes in spacetime, and are
known as {\it orientifold planes}.

The boundary conditions on the open-string end-points are determined
by the precise element(s) of $\Gamma$ involving
orientation-reversal. Generically the open strings turn out to have
Neumann (N) boundary conditions in some directions, and Dirichlet (D)
boundary conditions in others. The latter conditions prevent the
string end-points from moving in some directions -- thus the string
appears to be ``nailed'' to a hypersurface embedded in spacetime. This
hypersurface is called a {\it D $p$-brane}, where $p$ is the
number of {\it spatial} dimensions that it spans.

Thus an orientifold model contains two kinds of ``topological
defects'' within it: orientifold planes and D-branes. The physics of
these two types of objects is rather different at first sight. Planes
are supposed to be static objects, while it was proposed some time ago
that D-branes are dynamical\refs{\dlp}. Indeed, D-branes typically
carry integer charge with respect to some Ramond-Ramond (RR) gauge
potential, and they can be put into correspondence with charged BPS
solitonic branes that were earlier discovered as solutions to the
classical equations of motion of the low-energy effective action (see 
Ref.\refs{\dufflu} and references therein).

Planes also turn out to carry RR charges, but they were not originally
associated to dynamical objects. Subsequent developments have shown
that there are certain similarities between D-branes and orientifold
planes, but that these are observable in regimes which are beyond the
reach of perturbation theory.

The special feature of D-branes is that instead of treating them as
classical soliton solutions, it is possible to study them through
open-string perturbation theory. This can be understood in the context
of a more general phenomenon.  Many kinds of solitonic branes (stable,
extended classical solutions) exist in string theory. Certain
couplings in the underlying theory make it possible for specific open
branes to terminate on other branes\refs{\stromopen, \townopen}. The
crucial point is charge conservation: this ``termination'' can occur
without destroying the charge carried by the open brane, thanks to the
presence of certain world-volume couplings.  From this point of view,
the D-branes are special only because the perturbative open string can
end on them.

There are branes on which perturbative open strings cannot end, but on
which the so-called $(p,q)$ strings\refs{\schwarzpq} can end. These
strings are related to the fundamental string by $S$-duality. On other
branes, no string can end but membranes are allowed to end. In this
sense, every brane is a ``D-object'' for some other species of open
brane. This generalization of the ``D-concept'' is one of the more
interesting developments in this area in the last two years, and we
will say more about it in due course.

Orientifolds based on the group $Z_2$ are particularly simple to
define, in every number of uncompactified spacetime dimensions $d\le
10$. In recent times they have been shown to display a remarkable
range of properties which vary in a very interesting way with
spacetime dimension, making each case a new problem to address rather
than just a repetition of the previous one. In the next section I will
provide a list of all the cases and describe how their ``personality''
varies from one dimension to the next. Some of the more interesting
personalities will be dealt with in more detail.

\newsec{The Unique Personality of Each Dimension}

We start with the type IIA/B string in 10 spacetime
dimensions. Compactify on an $n$-dimensional torus, and let $I^{(n)}$
denote inversion of all the torus coordinates. Then, the action of
$\Omega\cdot I^{(n)}$ is a symmetry of the theory if $n$ is even for
type IIB or odd for type IIA. Taking the quotient with respect to this
symmetry, one ends up with a theory that has 16, rather than 32,
supersymmetries, and lives in $d=10-n$ noncompact spacetime
dimensions.  In every case the theory is non-chiral, and
non-perturbatively dual to the toroidally compactified heterotic
string.

In this class of models, there are clearly $2^n$ fixed orientifold
planes. These fill all of the noncompact spacetime, hence they are
$p$-planes, where $p=9-n$. Additionally, tadpole cancellation requires
the vacuum to contain 32 Dirichlet $p$-branes. Thus the geometry can
be visualized as follows.
\bigskip

\centerline{\epsfbox{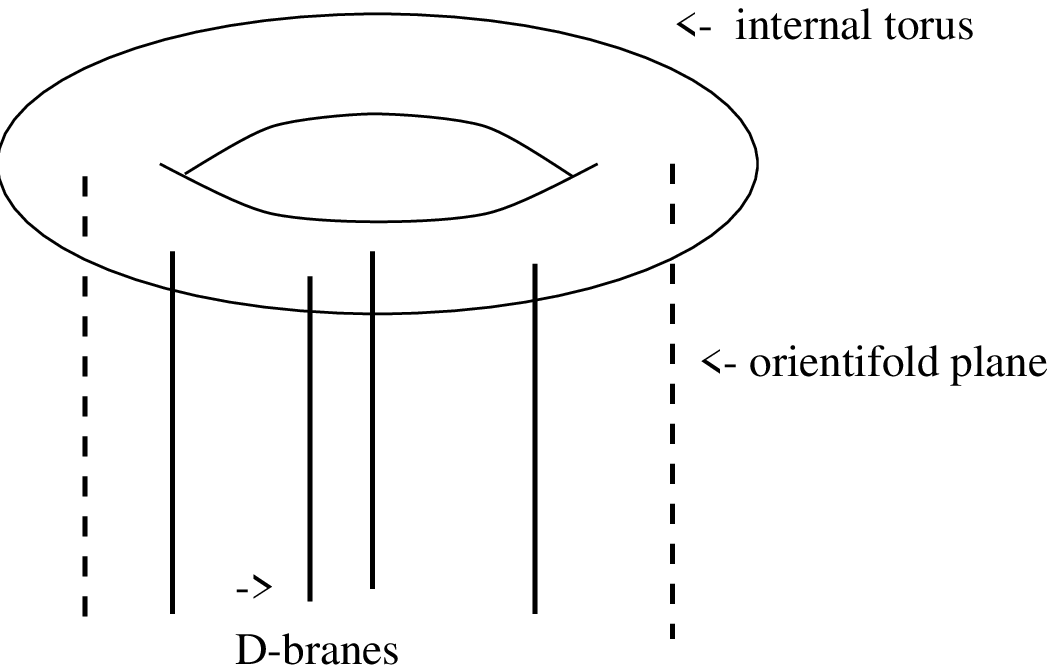}}
\bigskip

For a single D-brane the field theory on the world-volume is a $U(1)$
supersymmetric gauge theory. When $n$ parallel D-branes coincide, the
gauge symmetry is promoted, by new light open strings, to $U(n)$ with
adjoint hypermultiplets. If this coincidence takes place over an
orientifold plane, this is further promoted, by yet more light open
strings, to $SO(2n)$. Thus the $A$ and $D$ series of the simply-laced
Lie algebras are produced, but not the $E$-series. How exceptional
groups arise from branes is still a challenging and not fully
understood question. Some of the progress that has been made will be
reviewed below.

Note that since each D-brane moves in a pair with its mirror, the
number of branes at an orientifold plane (counting a brane and its
mirror as distinct) is conserved modulo 2. The ``standard''
configurations that have played the most important role in duality (so
far) are the ones with an even number of branes at each plane. In
these configurations, it is enough to concentrate on 16 of the 32
branes, since the other 16 move as their mirror images. Thus, for
example, well-separated branes in these models give a gauge group
$U(1)^{16}$ rather than $U(1)^{32}$.

However, there can be other components of moduli space, which, to my
knowledge, have not been explored in complete detail\foot{although
some related issues have been studied in the context of $Z_2\times
Z_2$ orientifolds in Ref.\refs{\bersix}.}. In these other components,
we encounter some ``half-branes'' stuck at fixed points, and the rank
of the gauge group coming from branes can be reduced. In what follows,
I will concentrate largely on the standard component of moduli space,
and mention the existence of the others only when relevant.

I now turn to the various interesting dynamics manifested by each of
these models for $1\le n\le 9$, corresponding to uncompactified
dimensions $d$ in the range $9\ge d \ge 1$.
\bigskip

\centerline{
{\bf\underbar{d = 9} } 
}

In this case we have compactified on a circle of radius $R$, whose
coordinate is labelled $x^9$. There are two 8-planes at $x^9=0,\half
R$ and sixteen 8-branes at arbitrary values of $x_9$ between 0 and
$\half R$.  As mentioned above, on the remaining segment $\half R\le
x^9 \le R$, there are sixteen ``mirror'' branes.

The 8-branes each carry unit charge with respect to the 9-form
potential that is present in the type IIA string. Charge cancellation
on the compact internal space then requires that the two 8-planes each
carry $-8$ units of charge. For general locations of the 8-branes,
there will be charge gradients, or electric fields, in the model. This
means, in particular, that the effective theory in the bulk region
between the branes and planes is {\it massive} type IIA
supergravity\refs{\romans,\pold}. The fields vanish only if precisely
eight 8-branes (along with their mirrors) are located on top of each
of the two 8-planes. The gauge group in that case is $SO(16)\times
SO(16)$.

For configurations which do not cancel charge, it has been noted that
the electric fields are accompanied by dilaton gradients over the
compact space\refs{\polwit}. One consequence is that the theory is
strongly coupled in many regions of the moduli space that is
parametrized by the radius and the brane locations. Indeed, in a
configuration of $n_1$ 8-branes at one 8-plane and $n_2$ 8-branes at
the other, the coupling diverges at the radius
\eqn\diverg{ 
R = \half\sqrt{8 - min(n_1,n_2) }}
In particular, this avoids a potential conflict with the prediction of
type I/heterotic duality. Subsequently the strongly coupled regions
have been studied, and it has been argued that at strong coupling, the
orientifold planes can ``emit'' upto two extra 8-branes which give
rise to enhanced gauge symmetries, even including $E_n$
factors\refs{\morsei}.

But the most fascinating phenomenon exhibited by this model becomes
evident in a certain limit. In the strong coupling limit, the
10-dimensional type IIA string acquires an extra space dimension and
becomes 11-dimensional ``M-theory''\refs{\townm,\witvar}. In a similar
fashion, the orientifold of type IIA on $S^1/Z_2$, in the
corresponding limit, also acquires an extra dimension and becomes an
``orientifold'' of M-theory on $S_1/Z_2$. This theory has been argued
to be dual to the $E_8\times E_8$ heterotic string\refs{\horwit}. It
is important that the limit should be taken starting from the
configuration in $d=9$ where the charges are locally cancelled, so
there are eight 8-branes, along with their mirrors, at each
orientifold plane. The gauge symmetry of this configuration,
$SO(16)\times SO(16)$, becomes enhanced to $E_8\times E_8$ in the
strong coupling limit. It is noteworthy that the limiting theory
is {\it chiral}.

\bigskip

\centerline{
{\bf\underbar{d = 8} } 
}

This is the $T^2/Z_2$ orientifold of the type IIB string, with 4
orientifold 7-planes. These sit at the points $(0,0), (0,\half R_2),
(\half R_1, 0), (\half R_1, \half R_2)$ on the 2-torus (which we have
taken to be square and whose sides have length $R_1,R_2$). In addition
there are 16 7-branes, and their images, at arbitrary points of the
2-torus.

The 7-branes and 7-planes are electrically charged under the dual
8-form gauge potential $C^{(8)}$ satisfying ${}^* d{\tphi} = d
C^{(8)}$ where ${\tphi}$ is the Ramond-Ramond scalar of the type IIB
string. Alternatively one can think of them as magnetically charged
with respect to $\tphi$ itself. The charges are 1 and $-4$
respectively for branes and planes. In the charge-cancelling
configuration, four 7-branes sit on every 7-plane and the gauge
symmetry is $SO(8)^4$.

In configurations which do not cancel charge, there turns out to be a
serious problem with the orientifold description\refs{\senf}. The
gradients this time are those of the field $\tphi$. However, the type
IIB string has an $SL(2,Z)$ S-duality which acts on the complex scalar
$\tau = \tphi + ie^{-\phi}$ by projective transformations. Thus a
gradient of $\tphi$ actually implies a (holomorphic) variation of
$\tau$ with the 2-torus coordinate $z$. It turns out that in the
orientifold description the complex scalar $\tau$ acquires a {\it
negative} imaginary part for some $z$, as soon as field gradients
arise. This is a breakdown of the description itself, since it makes
$\phi$ undefined.

Luckily, there is an alternative description of the model which, in a
certain limit, reduces to the orientifold description at the
charge-cancelling configuration, but is believed to correctly describe
the model more generally. In this description, called
``F-theory''\refs{\vafaf}, there are no orientifold planes at all. It
is a compactification of type IIB on a base $P_1$, where $\tau$ varies
over the $P_1$ base, describing an elliptic fibration. There are 24
7-branes corresponding to singularities of the fibration.

Not all of these 7-branes are D-branes. On separating the 24 7-branes
into four groups of six coincident 7-branes each, one recovers the
orientifold model in the charge-cancelling configuration. It follows
that one should identify four of the six 7-branes in a group with
D-branes, while the other two, when they coincide, behave like an
orientifold plane. Conversely, the non-perturbative description of the
orientifold theory requires each plane to ``split'' into two
7-branes. These are magnetically charged, not with respect to $\tphi$
but under an $SL(2,Z)$ transform of $\tphi$. The orientifold limit
with gauge group $SO(8)^4$ is a special case in which $\tau$ remains
constant over the base but the fibre acquires orbifold singularities
at 4 points. The base itself is an orbifold limit, $T^2/Z_2$, of
$P_1$\refs{\senf}. 

The splitting of orientifold planes into 7-branes suggests that
orientifold planes too must have some dynamics. Indeed, although they
cannot carry perturbative gauge fields on their world-volume, planes
can, and do, carry gravitational couplings, as argued recently in
Ref.\refs{\djm}. The predicted couplings in the present case turn out
to be consistent with the nonperturbative splitting phenomenon.

In this example, the occurrence of enhanced gauge symmetry (including
exceptional factors) involves interesting physics from the brane point
of view. It has been noted\refs{\dmf} that besides the
orientifold model above, one can consider other limits of F-theory
compactification in which $\tau$ is constant over the base. Remarkably
in these other limits, the base need not be an orbifold limit of
$P_1$, but the fibre is constrained to have a fixed complex-structure
modulus $\tau$. There are actually two such branches, with $\tau=i$
(branch I) and $\tau = {\rm exp}(2\pi i/3)$ (branch II), both of which
are fixed points of $SL(2,Z)$.

On these branches, one finds that F-theory branes move in groups of 3
(for branch I) or in groups of 2 (for branch II). Along these
branches, from the F-theory point of view one can show that
exceptional gauge symmetries arise: $E_7$ on branch I and $E_6$, $E_8$
on branch II. These branches contain in particular the orbifold limits
of K3. Branch I contains the $T^4/Z_4$ limit of K3 on it, while branch
II has the $T^4/Z_3$ and $T^4/Z_6$ limits. Except for the fact that
$\tau$ is at a strongly-coupled value (i.e. of order 1), these
branches look very much like orientifold compactifications. Thus one
is led to conjecture\refs{\dmf} the existence of a generalized
``non-perturbative orientifold theory'' corresponding to these
branches.

Recently, some progress has been made in identifying the origin of
these exceptional symmetries, in terms of branes on which open $(p,q)$
strings end\refs{\johan,\gabz}. In particular, the existence of
multi-pronged strings carrying different $(p,q)$ charges at the
various end-points has been argued\refs{\gabz} to play a key role in
the phenomenon. Such string configurations exemplify the possible
generalized D-objects that could play a role in a full nonperturbative
description of string theory.

Our earlier discussion on the existence of disconnected branches in
the moduli space becomes relevant for the first time in this
example. At each of the four orientifold planes, one can start by
placing an {\it odd} number of branes, for example $7,7,9,9$. Since
branes can move only in mirror pairs, there will always be a net brane
at each fixed plane. The enhanced symmetry corresponds to
non-simply-laced groups, for example in this case we get $SO(7)^2
\times SO(9)^2$. The total rank of the gauge group from the
open-string sector is reduced from 16 to 14. The relationship of this
branch to CHL compactifications\foot{suggested in a different
context in Ref.\refs{\witfive}} and to F-theory remains to be worked
out.
\bigskip

\centerline{
{\bf\underbar{d = 7} } 
}

This is a $T^3/Z_2$ orientifold of the type IIA string. It corresponds
to a theory in 7 spacetime dimensions, with eight 6-planes and sixteen
6-branes.  These objects carry magnetic charge for a 7-form potential,
dual to the RR 1-form of type IIA. The branes as always carry charge
$+1$, and the planes this time have charge $-2$. Thus the
charge-cancelling configuration requires two branes on each plane,
leading to the gauge group $SO(4)^8$.

It is known that the strong coupling limit of type IIA
string theory is ``M-theory'', and this fact permits a transparent
description of the dynamics of this vacuum configuration.
From the M-theory point of view, the RR 1-form of type IIA is actually
a Kaluza-Klein gauge field. Objects carrying magnetic charge with
respect to a KK gauge field are called Kaluza-Klein monopoles. The
6-branes indeed arise as such monopoles, from the solitonic point of
view. It is less clear a priori how the orientifold planes, which also
carry charge, should be understood as KK monopoles, but we will see
that in some sense they can be thought of in this way.

Since M-theory compactified on a circle is the type IIA string, one
might try to think of this orientifold as M-theory on a circle
multiplied by $T^3/Z_2$. A more careful analysis\refs{\seibthree}
reveals that the $Z_2$ acts on the circle factor as well, so the model
is actually defined as M-theory on $T^4/Z_2$, an orbifold limit of the
4-manifold K3. It has been argued that taking into account all
non-perturbative effects, the full moduli space of the theory is the
moduli space of the general K3 manifold.

The charge-cancelling configuration is simply a degenerate K3 with 8
$D_2$ singularities. Some more details are known about this
model\refs{\seiwitthree}. If we concentrate on an isolated 6-brane far
away from other branes and planes, its moduli space looks like the
noncompact Taub-NUT 4-manifold. Similarly, considering an isolated
6-plane away from everything else, the moduli space looks like a
closely related but distinct manifold: the Atiyah-Hitchin space, which
arises as the moduli space of two-monopole solutions in Yang-Mills
theory.

An interesting consequence can be derived from these properties. We
have already encountered the fact\refs{\djm} that orientifold planes
carry definite WZ gravitational couplings localized on their
world-volume. Using the observations in the previous paragraph, which
define the transverse space to 6-branes and 6-planes in the present
model, Sen\refs{\sensix} has obtained the coefficient of the
$C^{(3)}\wedge p_1(R)$ WZ coupling on both these objects, in agreement
with the prediction of Ref.\refs{\djm}.
\bigskip

\centerline{
{\bf\underbar{d = 6} } 
}

This is an orientifold of type IIB on $T^4/Z_2$ with 16 5-branes and
16 5-planes in the vacuum. The branes and planes carry equal and
opposite magnetic charge under the RR 2-form potential of the type IIB
string. The charge cancelling configuration simply requires one
5-brane on top of every 5-plane. The gauge group is Abelian, namely
$U(1)^{16}$, in this case.

This model has been argued\refs{\bsv} to be dual to a conventional
{\it orbifold} of type IIA on $T^4/Z_2$. Nonperturbatively, its moduli
space coincides with that of IIA on K3. The reason for this
correspondence is less mysterious when one realizes that we have
compactified the $n=3$ model on a circle and T-dualized. Hence we
should expect to find M-theory on $S_1\times K3$, which is the same as
type IIA on K3. Note that the set of models with $n=2,3,4$ form a
chain which corresponds to compactifications of F-theory, M-theory and
IIA string theory on the same manifold, namely K3. Beyond $n=4$
something different happens.
\bigskip

\centerline{
{\bf\underbar{d = 5} } 
}

This theory is the $T^5/Z_2$ orientifold of type IIA, with sixteen
4-branes and $2^5=32$ orientifold 4-planes. For the first time, the
planes have fractional magnetic charge, $-\half$, with respect to the
3-form potential $C^{(3)}$ of the type IIA string. Thus there is no
charge-cancelling configuration in the usual component of moduli
space. However, there is a different component of moduli space on
which we can choose the 32 4-branes to lie one on each of the 32
planes\refs{\witfive}. Since away from the plane the branes have to
move in pairs, it follows that starting from this configuration the
branes cannot move any more! Since $N=2$ supermultiplets in 6d have a
gauge field together with four scalars, the absence of moduli for the
branes must mean that the gauge fields have been projected out as
well, leading apparently to no gauge group from the branes. That makes
this a rather intriguing case.

Returning to the usual component of moduli space, we note that this is
the usual Narain moduli space for compactification of the heterotic
string to 5 spacetime dimensions, namely $SO(5,21)$ modded out by the
standard continuous and discrete symmetries. One could again ask what
happens if we take the ten-dimensional type IIA coupling to be large,
so that the type IIA theory goes over to M-theory. If such a limit
exists for the present compactification, then it should be
6-dimensional and retain the $SO(5,21)$ moduli space. Remarkably there
is a known candidate for a dual to this theory: the K3
compactification of the type IIB string. This leads one to
conjecture\refs{\dmm,\witfive} that there is a $T^5/Z_2$ orientifold
of M-theory, and that it is dual to type IIB on K3.

Unlike all the $Z_2$ string orientifolds (in $D<10$), but similarly to
the Horava-Witten model, this M-theory orientifold is {\it
chiral}. Instead of D-branes, it has 32 M-theory 5-branes in the
vacuum (again, 16 independent 5-branes and their mirrors). It has
potential gravitational anomalies, which are cancelled both globally
(by summing up bulk and brane-worldvolume contributions) and locally
(by anomaly inflow from the bulk onto the 5-branes and the orientifold
5-planes).

Thus the $D=5$ string orientifold exhibits a totally different
personality from the cases of $d=8,7,6$ (recall that those formed a
chain). It ``lifts'' to a chiral theory which cannot be described in
string perturbation theory. M-theory 5-branes are not D-branes but
rather generalized D-objects, on which M-theory 2-branes can
end\refs{\stromopen,\townopen}.

As noted above, this orientifold theory (and the following ones in
dimensions $d< 5$) have orientifold planes with fractional
charges. This would appear to contradict Dirac quantization: one can
imagine taking a suitable D-brane around the fractionally charged
plane and observing a nontrivial monodromy. However, it turns
out\refs{\witflux,\djm} that the quantization rule for $2\le d \le 5$
gets modified as a consequence of anomalous Bianchi identities in
$6\le d \le 9$. The latter arise due to fractional contributions to
the curvature invariants coming from each orientifold plane. This
explains the apparent puzzle: though the naively defined flux is
fractional, the gauge-invariant flux that is observable is not.
\bigskip

\centerline{
{\bf\underbar{d = 4} } 
}

Starting from this case, which is type IIB on $T^6/Z_2$, a new chain
appears. As an orientifold, this has 32 3-branes and 64 orientifold
3-planes filling the noncompact space. The planes have fractional
charge $-{1\over 4}$ with respect to the self-dual 4-form $C^{(4)+}$ of
the type IIB string. Very much like the $D=8$ case described above,
this model can be described as the limit of an F-theory
compactification on the manifold $T^8/Z_2$. The main difference is
that $T^8/Z_2$ is not the limit of a suitable smooth 8-manifold, as it
has terminal singularities.

In terms of F-theory compactification, one understands the existence
of the 32 3-branes (or rather 16 independent 3-branes if we don't
count the mirrors) as follows. It has been observed\refs{\svw} that
F-theory compactified on 8-folds to 4 spacetime dimensions has a
potential tadpole for $C^{(4)+}$ which is cancelled by introducing
$\chi/24$ 3-branes in the vacuum, where $\chi$ is the Euler
characteristic of the 8-fold. In the orbifold sense, it is easy to
compute that $T^8/Z_2$ has $\chi=384$, from which it follows that 16
3-branes are expected.

Unlike 7-branes, 3-branes are self-dual under $SL(2,Z)$. This fact
allows one to deduce some interesting properties of WZ couplings on
3-branes and 3-planes from this model\refs{\djm}.
\bigskip

\centerline{
{\bf\underbar{d = 3} } 
}

This is the $T^7/Z_2$ orientifold of type IIA, which by arguments
similar to those in the $d=7$ case, can be mapped to the
compactification of M-theory on $T^8/Z_2$. Again a tadpole is known to
arise, this time of the M-theory 3-form $C^{(3)}$, for general
M-theory compactifications on 8-folds, and this is cancelled by
condensing 16 M-theory 2-branes in the vacuum, consistent with the
structure of the orientifold model.

The orientifold 2-planes have electric charge $-{1\over 8}$ with
respect to the M-theory 3-form potential.
\bigskip

\centerline{
{\bf\underbar{d = 2} } 
}

This model is a type IIB orientifold on $T^8/Z_2$. It is equivalent to
the orbifold compactification of type IIA on $T^8/Z_2$, just as the
orientifold of IIB on $T^4/Z_2$ was equivalent to the orbifold of IIA
on $T^4/Z_2$, namely the K3 compactification of IIA.

This is the third and last in the series of orientifolds that are dual
to $T^8/Z_2$ compactifications of F-theory, M-theory and IIA. This
time, viewed as a IIB orientifold it has 16 independent type IIB D
1-branes in the vacuum. But as a IIA compactification, it needs 16
1-branes of type IIA to cancel the tadpole of the 2-form field. So
this compactification has fundamental strings condensed in the vacuum.
One has to remember that the concept of moduli space strictly does not
make sense in such low dimensions.
\bigskip

\centerline{
{\bf\underbar{d = 1} } 
}

The last theory that we consider is the $T^9/Z_2$ orientifold of the
type IIA string, to $0+1$ spacetime dimensions. In other words, all of
space has been compactified. This time, there are 16 independent D
0-branes in the vacuum. This case might be expected to be similar to
the ones in $d=5$ and 9, which lift to chiral M-theory orientifolds in 6
and 10 dimensions. But it is worth appreciating ``the little
differences''\refs{\qt}. 

Consider the M-theory orientifold on $T^9/Z_2$, down to 2 spacetime
dimensions\refs{\dmm}. In this model again there are potential
gravitational anomalies. Hence there must be twisted sectors. Recall
that in $d=5$ above, the twisted sector consisted of 4-branes, which
are known to go over in strong coupling to M-theory 5-branes. Hence in
M-theory on $T^5/Z_2$ the twisted sector consists of 16 5-branes,
which carry tensor multiplets on their world-volume.

In the present case we have 0-branes as the twisted sector. That seems
to present a paradox. In strong coupling, when the IIA string becomes
M-theory, and its D 4-branes become M-theory 5-branes, the D 0-branes
do not gain a dimension! Rather, they are just Kaluza-Klein momentum
modes of the higher dimensional theory. So M-theory on $T^9/Z_2$ will
not have branes in the twisted sector, but something else.

Whatever else appears, must serve to cancel the gravitational anomaly
in $1+1$ dimensions coming from the untwisted sector. One way to see
what these new states are\refs{\djm} is to observe that D 0-branes are
BPS and are annihilated by the RHS of the supersymmetry algebra
\eqn\zerodimalg{
\{Q_i,Q_j\} = (P_0 + Z)\delta_{ij}}
where $Z$ is the central charge and $P$ is the energy. In one higher
dimension, the analogue of this algebra is the {\it chiral}
supersymmetry algebra obtained by replacing $Z\rightarrow P_1$, thus
\eqn\onedimalg{
\{Q_i,Q_j\} = P^+ \delta_{ij}}
where $P^+\equiv P_0 + P_1$. Given that D 0-branes are BPS states of
the $0+1$d algebra, they must go over to supersymmetric states of the
$1+1$d theory which are annihilated by $P^+$. This in turn is just
the Dirac equation, and the states we are looking for are
therefore {\it chiral fermions} in $1+1$d, which are supersymmetry
singlets.  

Indeed, an analysis of gravitational anomalies in the $1+1$d theory
obtained by compactifying M-theory on $T^9/Z_2$ reveals that 512
chiral fermions are needed for anomaly cancellation. But that is
exactly equal to the number of fixed points of the $Z_2$ action on
$T^9$, namely $2^9$. So it is consistent to assume one chiral fermion
per fixed point\refs{\dmm}. This model has been argued to be dual to
type IIB on $T^8/Z_2$, which is also chiral and
anomaly-free\refs{\dmn}. 

Thus we see that gravitational anomaly cancellation in M-theory on
$T^9/Z_2$ takes a rather different route from the $T^5/Z_2$ case.

\newsec{Conclusions}

The fact that such a simple class of superstring compactifications
displays such highly dimension-dependent personalities is most
intriguing. There are other situations in which the personalities of
various dimensions are manifest: for example, the U-dualities of
toroidally compactified strings, and the dynamics of supersymmetric
gauge theory with 8 supersymmetries in spacetime dimensions $2\le d\le
6$.

Admittedly the dimension dependence has a lot to do with
supersymmetry, whose representation theory has long been known to vary
in a beautiful way with dimension\refs{\salamsez}. And yet the
phenomena that we have discussed above are not merely classical or
representation-theoretic, but involve the full quantum dynamics of
string theory, despite the absence to date of a concrete
nonperturbative formulation.  This is, to my mind, highly
encouraging. It suggests that in a suitable framework yet to be
discovered, the special value of spacetime dimension in which we live
will be selected in a dynamical fashion.

\newsec{Acknowledgements}

I am grateful to Keshav Dasgupta and Dileep Jatkar for enjoyable
collaborations, and to Atish Dabholkar, Sumit Das, Avinash Dhar,
Gautam Mandal and Ashoke Sen for helpful discussions on the issues
discussed above. I also thank Chris Hull for encouraging me to write
up these observations in the present form.

\listrefs
\end